\documentclass[preprint]{revtex4}
\usepackage{natbib}
\usepackage{amsmath}
\usepackage{graphicx}
\def\beq{\begin{eqnarray}}
\def\eeq{\end{eqnarray}}

\usepackage{amsbsy}
\usepackage{epsf}
\def\X{{\mathrm{x}}}
\def\Y{{\mathrm{y}}}

\def\sf{{\mathrm{S}}}

\def\k{{\rm k}}
\def\n{{\rm n}}
\def\p{{\rm p}}
\def\e{{\rm e}}
\def\l{\Lambda}
\def\c{{\rm c}}
\def\h{\Sigma}
\def\s{\mathrm{s}}

\newcommand{\be}{\begin{equation}}
\newcommand{\ee}{\end{equation}}
\newcommand{\bear}{\begin{eqnarray}}
\newcommand{\eear}{\end{eqnarray}}

\begin{document}

\title{The dynamics of dissipative multi-fluid neutron star cores}

\author{B. Haskell$^{1}$, N. Andersson$^2$, G.L. Comer$^3$}
%{B. Haskell, N.Andersson \\
%School of Mathematics, University of Southampton,
%Southampton, SO17 1BJ, United Kingdom\\}
\affiliation{
$^1$ Astronomical Institute ``Anton Pannekoek'', University of Amsterdam, Science Park 904, 1098 XH Amsterdam, Netherlands\\
$^2$School of Mathematics, University of Southampton, SO17 1BJ Southampton, UK\\
$^3$Department of Physics and Center for Fluids at All Scales, Saint Louis University, St Louis, Missouri 63103, USA }

%\maketitle

\begin{abstract}
We present a Newtonian multi-fluid formalism for superfluid neutron star cores, focussing  on the  additional dissipative terms that arise when one takes
into account the individual dynamical degrees of freedom associated with the coupled ``fluids''. The problem is of direct astrophysical interest as
 the  nature of the dissipative terms can have significant impact on the damping of the various oscillation modes of the star and the associated
gravitational-wave signatures. A particularly interesting application concerns the gravitational-wave driven instability of  f- and  r-modes.
We apply the developed formalism to two specific three-fluid systems:
(i)  a hyperon core in which both $\Lambda$ and $\Sigma^{-}$ hyperons are present, and (ii) a core of deconfined quarks in the colour-flavour-locked phase in which a population of neutral $K^{0}$ kaons is present. The formalism is, however, general and can be applied  to  other problems in neutron-star dynamics (such as the effect of thermal excitations close to the superfluid transition temperature) as well as laboratory multi-fluid systems.

\end{abstract}

\maketitle

\section{Introduction}

Neutron stars provide unique laboratories for the study of the state of matter under extreme conditions. Comprising roughly one and a half solar masses within a radius
of about 10 km, these very compact objects are likely to have core densities reaching several times the nuclear saturation density. Thus, they offer the opportunity to probe the cold, high density region of the QCD phase diagram (that cannot be explored in laboratory experiments).  Improvements of our theoretical description of these systems are crucial. Not only is a wealth of data from  X-ray and radio observations available already, but ground-based gravitational-wave detectors have  reached their initial design sensitivities and are now being upgraded to the
second generation level \cite{ligo}. This is  significant since gravitational-wave observations would provide truly complementary information about these exotic objects \cite{etpaper}. Gravitational waves carry an imprint of the dynamics of the internal, high density, regions of the star while
the electromagnetic signature relies on a complex interaction with, and processes within, the star's magnetosphere.

Accurate modelling of neutron star interiors is important, but fraught with difficulty. A detailed model of neutron star dynamics must
not only account for exotic states of matter at high density. It must also consider the interaction of the crustal nuclei (the outer kilometer or so)
with the star's core, the presence of a strong magnetic field and various superfluid/superconducting states.
Superfluidity, in particular, adds dimensions to the problem by introducing new dynamical degrees of freedom.
This can  have profound consequences for the dissipation in the interior of the star \citep{EP,monster}.
In the last few years, the effect of superfluidity on the damping of the gravitational-wave driven instability of the so-called r-modes has sparked much  interest (see  \cite{sfr1,rmode,hyperon} for recent discussion),
and we now know that the details of the interior microphysics can play a significant role. Effects such as the vortex-mediated mutual friction damping 
\cite{LM,YL,rmode} and the enhanced bulk viscosity due to the presence of hyperons \citep{LO,nayyar, Ghyp,rmode} or of a quark condensate in the colour-flavour locked (CFL) phase \citep{cfl} 
can all modify the damping of the unstable modes considerably. 
It has, in fact,  recently been shown that measured  neutron star spins and inferred
temperatures in Low-mass X-ray binaries are not consistent with the standard models for the 
r-mode instability window. These results suggest that 
some form of enhanced damping is  required \cite{hoprl,degen}. Further theoretical 
investigation is clearly necessary if we want to be able to make direct, quantitative, 
comparisons with the X-ray data. 

The aim of this paper is to construct a general framework for dissipative multi-fluid neutron star cores, building on
 the variational formulation of two-fluid neutron star hydrodynamics \citep{Prix, monster} (see also \cite{chamel1,chamel2}).  We will focus on the dynamics of three-fluid systems at finite temperature.
This problem is  relevant both in the case of hyperon cores, where one expects not only neutrons and protons, but also a population of $\Lambda$ and $\Sigma^{-}$ hyperons \citep{Glenn}, and in the case of cores of deconfined quarks in the CFL phase, where one should account not only for the quark condensate but also for the thermal excitations of the system and possibly  a population of kaons \cite{reviewA,alf1,kaon1}.
Moreover, given the strong density dependence of the various superfluid pairing gaps \cite{nuclpa} there will \underline{always} be regions in the star that are
close to the critical superfluid transition temperature. In these regions thermal effects are not negligible. Again, this situation can be
described by a three-fluid system comprised of neutrons, protons and thermal excitations \cite{thermal}.

We pay particular attention to the dissipative terms in the hydrodynamical equations. It is well known from the study of superfluid $^4$He that there are more dissipation coefficients in a superfluid system than in the standard Navier-Stokes description. However, although this was first pointed out in \cite{EP} in the context of
neutron star physics, and several authors have studied the effect of superfluidity on the standard dissipation coefficients (see \cite{visc1, visc} for recent examples), 
very little effort has been made to understand
the nature (and dynamical role) of these extra coefficients. Having said that,  there have been  some recent attempts to quantify the effect of
superfluidity on dissipative neutron star oscillations \citep{kant, hyperon}.
In this paper we  build on the work of \cite{monster}, correcting some conceptual issues, in order to develop a general formalism for dissipation in multi-fluid systems.
We are developing the formalism with neutron star problems in mind, but the framework is sufficiently
general that it can be applied to entirely different systems, e.g. ones that can be studied in the laboratory. In fact, a three-fluid model is required to  account for the interactions between phonons and rotons (and the associated thermal conductivity) in $^4$He \citep{kal,helium}. A recent discussion of extended thermodynamics and causal 
heat conduction \cite{heat} in relativistic models should also be noted. Finally, it is worth pointing out the close connection between the multi-fluid model that we advocate and the general framework of extended irreversible thermodynamics \cite{extro}.

The paper is structured as follows: In section II we discuss the key length-scales associated with, and the general notion of, a
``multi-fluid'' system. In section III we outline our flux-conservative multi-fluid formalism, leading to the general form for the dissipation coefficients presented
in section IV. In section V we discuss the irrotational constraint for superfluid flows. Section VI then deals with perturbations of the multi-fluid system for backgrounds representing slowly rotating stars, while section VII is devoted to two explicit examples, (i) that of a hyperon core and (ii) that of a core of deconfined quarks in the CFL-K$^0$ phase. Finally, we present our conclusions in section VIII.

Throughout the paper we use a coordinate basis to represent tensorial relations. We thus distinguish between contravariant (of the form $v^i$) and covariant (of the form $v_i$) objects, and raise and lower indices with the (three dimensional flat-space) metric $g_{ij}$.
We shall also identify the different fluids with constituent indices, so that $v_\X^i$ will be the velocity of the x-th fluid, while $v_\Y^i$ relates to the y-th. We do not assume that the constituent indices have any geometric meaning (although it is worth noting that it is possible to develop the idea of ``chemical covariance''), so they can be placed either as sub- or superscripts, depending on which makes the relevant expression the clearest. 

\section{The different scales in the problem}

Before we discuss specific models, it is useful to consider the ``big picture'' of fluid dynamics
for ultra-dense matter. This is a key issue, because many neutron star scenarios require an understanding
of dynamics on the macroscopic scale, for which hydrodynamics is the natural tool. At the same time,
these models must build on an understanding of the microsphysics, for which we need to turn to nuclear
and particle physics.  A properly formulated
theory of fluid dynamics tracks the evolution of fluid elements --- their
trajectories through spacetime and how the various thermodynamic properties
(i.e.~an equation of state with relatively few parameters) change.  Each fluid
element is small enough that it can be considered as a ``point'', but large
enough to contain many particles so that a ``smooth-averaged'' thermodynamic description is
appropriate.

Let us consider some of the issues involved;  first of all,
 nuclear and particle physics models tend to assume global Lorentz
invariance for all (matter) fields.  This means that gravity is
neglected.  However, one thing that general relativity does well is break
global Lorentz invariance,  while maintaining it locally.  Glendenning \cite{GlennBook} 
points out that over the scale of a fluid element in a neutron star, the change in the spacetime metric
is negligible and one can always erect a local reference frame where Lorentz
invariance holds.  Basically, there is a clear separation of scales and one can  use Lorentz-invariant
nuclear and particle physics models {\em provided} one accepts that ``global''
in this context means ``on the scale of a fluid element''. In essence, for a compact star,
we do not have a global phase space to be applied to all
particles; rather, each fluid element has its own phase space that applies to
the particles within that element.

This has repercussions when we consider the micro-physics. For example,
in the case of fermions (like neutrons and protons) one may ask to what extent one can
ignore particle states ``below'' the Fermi surface. On the one hand, one might
argue ``yes'' because the lowest energy is zero, and particles occupying
that and the other energy states below the Fermi surface are
locked in, on average, because nearby states will be occupied, and the
available energy reservoir may not be sufficient to launch particles above
the Fermi surface.  In this line of reasoning, it will only be states near
the Fermi surface that participate in the fluid dynamics.  To some extent this argument is correct, and only particles near the
Fermi surface contribute to the transport coefficients required in the fluid model (viscosities, thermal conductivity etcetera).
However, at the same time \underline{all} particles contribute to the global dynamics, as represented by the star's large scale oscillation modes. 
This is easy to see if we consider the fact that these modes couple to the gravitational field, which is sourced by all the matter. 
The fluid dynamics model requires information about both bulk properties and transport phenomena, making its formulation  a challenge. 

Next, let us consider the scales associated with fluid dynamics. This problem is central
to the analysis in this paper. In order to discuss a ``multi-fluid'' system, we obviously need to have some
understanding of what this concept entails.
For ordinary matter, the scale is simply set by interparticle collisions.
Since we need to associate a single ``velocity'' with each fluid element, the particles
must be able to equilibrate in a meaningful sense (e.g. have a velocity distribution with a well defined peak).
The relevant length-scale is the mean-free path. This concept is closely related to the
shear viscosity of matter. In the case of neutrons (which dominate the outer core of a typical neutron star)
we would have
\be
\lambda \approx { \eta \over \rho v_F} \approx 10^{-4} \left( {\rho \over 10^{14}\ \mbox{g/cm}^3} \right)^{11/12} \left( {10^8 \ \mbox{K} \over T}\right)^2 \ \mbox{cm} \ ,
\ee
where $v_F$ is the relevant Fermi velocity and we have used the estimate for the neutron-neutron scattering shear viscosity $\eta$ from \citep{visc}.

This estimate gives us an idea of the smallest scale on which it makes sense to consider the
system as a fluid (about a micron). It also hints at systems with distinguishable multi-fluid behaviour. Consider a
system with two particle species, and assume that the mean-free path associated with  scattering of particles of the same kind
is (for some reason) significantly shorter than the scale for cross-species collisions. Then we have two clearly defined ``fluids''
and it makes sense to consider the problem using the machinery that we will discuss later. This is, however,
not quite the situation that is motivating the present work. Our focus is on systems that exhibit superfluidity.
At the most basic level, superfluidity implies that there is no friction impeding the flow. Technically, the mean-free path diverges
and the previous argument does not work anymore. However, a superfluid system has a different scale associated with it;
the so-called coherence length. The coherence length arises from the fact that a superfluid is a ``macroscopic'' quantum state,
the flow of which depends on the gradient of the phase of the wave-function (the so-called order parameter).
On some small scale, the superfluidity breaks down due to quantum fluctations. This scale is known as the coherence
length. It can be taken as the typical ``size'' of a Cooper pair in a Fermionic system.
On any larger scale the system exhibits fluid behaviour (in the sense of the Landau two-fluid model for Helium \citep{kal,helium}).
For neutron-star superfluids, the coherence length is of the order of tens of Fermi \citep{Mendella, supercon}; much much smaller than the mean-free path
in the normal fluid case. This means that superfluids can exhibit extremely small scale dynamics. Since a superfluid is inviscid,  superfluid neutrons and superconducting protons (say) do not scatter (at least not as long as thermal excitations can be ignored) and hence the outer core of a neutron star requires a multi-fluid treatment. 

It would seem then, that one can meaningfully take
the fluid elements to have a size of the order of the coherence length.
However, in reality yet another length-scale needs to be considered. On scales larger than the Debye screening length, the electrons will be electromagnetically locked to the protons, forming a charge-neutral conglomerate that
\underline{does} exhibit friction (due to electron-electron scattering). Moreover, at finite temperatures we need to consider thermal excitations for both neutrons and protons, making the problem rather complex.

Furthermore, ideal superfluids are irrotational and neutron stars are not. In order to mimic bulk rotation the neutron
superfluid must form a dense array of vortices (breaking the superfluidity locally). This brings another length-scale
into the picture. In order to develop a useful fluid model, we need to average over the vortices as well. This makes the
effective fluid elements much larger. The typical vortex spacing in a neutron star is of the order \citep{supercon}
\be
d_\n \approx 4\times10^{-4} \left({P \over 1\ \mbox{ms}} \right)^{1/2} \ \mbox{cm} \ ,
\ee
where $P$ is the neutron star spin period.
For a slowly rotating object the fluid elements we consider (at the end of the day) may be quite large (although obviously still much smaller than the size of the star).

\section{Flux-conservative multi-fluid model}

We  take as our starting point the flux-conservative multi-fluid formalism developed in \cite{monster}.
This model combines the  conservation laws for mass, energy and angular momentum with the variational approach  developed by \citet{Prix} (see also \cite{chamel1,chamel2}).
 In this model (which represents the Newtonian limit of Carter's convective variational formalism in relativity  \cite{carter,livrev})
the main variables are the particle fluxes $n_i^\X=n^\X v^\X_i$ (where $n^\X$ is the particle number density of the x-th component, and $v^\X_i$ its velocity) and the equations of motion are derived from a Lagrangian density $\mathcal{L}$ of the form
\be
\mathcal{L}=\sum_\X \frac{m^{\X}}{2 n_\X}  g_{ij}n_\X^in_\X^j-\mathcal{E}(n_\X,n_\X^i) \ .
\ee
This allows us to define the conjugate momenta
\be
p^i_\X\equiv\left.\frac{\partial\mathcal{L}}{\partial n^i_\X}\right|_{n_\X}=g_{ij}m^\X v_\X^j-\left.\frac{\partial\mathcal{E}}{\partial n^i_\X}\right|_{n_\X} \ ,
\ee
where we still need to provide the energy density functional $\mathcal{E}$, which includes the internal energy. 
Following \cite{monster} we  consider an energy functional that is manifestly isotropic and Galileian invariant, i.e. we take
\be
\mathcal{E}=\mathcal{E}\left(n_\X,w_{\X\Y}^2\right),
\ee
with the relative velocities defined by
\be
w_{\X\Y}^i=v_\X^i-v_\Y^i\;\;\;\;\;\mbox{and}\;\;\;\;\; w_{\X\Y}^2=g_{ij}w_{\X\Y}^i w_{\X\Y}^j \ .
\ee
The chemical potentials are then obtained from
\be
\mu^{\X}\equiv \left.\frac{\partial \mathcal{E}}{\partial n_\X}\right|_{w_{\X\Y}^2} \ ,
\ee
while the entrainment coefficients are given by
\be
\alpha^{\X\Y}\equiv \left.\frac{\partial \mathcal{E}}{\partial w_{\X\Y}^2}\right|_{n_{\X}} \ .\label{ent}
\ee
These coefficients describe the fact that the momentum of each species is not necessarily parallel to the associated flux. Instead, it takes the form:
\be
\pi^\X_i=n^\X p_i^\X=m_\X n^\X v_i^\X + 2 \sum_{\X \neq \Y}\alpha_{\Y\X} w_i^{\X\Y} \ .
\ee
The most common context in which this effect has been studied relates to neutrons and protons in neutron star cores. In this case the effect is due to the fact that, because of the strong nuclear interaction, each neutron is associated with a virtual cloud of protons. This modifies the effective neutron mass in a dynamical setting
\citep{Prix, cj}. Recently, the entrainment parameters in a hyperon core have also been discussed \cite{Gusakov1,Gusakov2}.
The usefulness of the entrainment concept for entropy, and its relation with thermal relaxation and causal heat conduction has also been explored in recent work \cite{heat, cesar,cesar2,mhd}

The entrainment is a dynamical  effect that arises naturally within the variational model.
In a practical application, it  depends on the nature of the microphysics that one includes in the equation of state (represented by $\mathcal{E}$).
As discussed in \cite{monster} the momentum equations take the form:
\be
\partial_t\pi_i^\X+\nabla_j\left(v_\X^j\pi_i^\X+D^{\X j}_i\right)+n_\X\nabla_i\left(\mu_\X-\frac{1}{2}m_\X v_\X^2\right)+\rho_\X\nabla_i\Phi+\pi^\X_j\nabla_iv_\X^j=f^\X_i \ ,
\label{euler}
\ee
where the tensor $D^\X_{ij}$ represents the viscous stresses, while the ``forces'' $f^\X_i$ allow for the transfer of momentum between the two components.
The particle mass densities are $\rho_\X$ and $\Phi$ is the gravitational potential. The latter satisfies the usual Poisson equation
\be
\nabla^2\Phi=4\pi G\sum_\X \rho_\X \ .
\ee
In the following we shall consider the case of an isolated system, for which  $\sum_\X f^\X_i=0$.

The continuity equations can be written
\be
\partial_t n_\X+\nabla_j\left(n_\X v_\X^j\right)=\Gamma_\X.
\label{continuity}
\ee
We will  assume overall mass conservation for our system, which in terms of the particle creation rates $\Gamma_\X$ leads to
\be
\sum_\X m_\X \Gamma_\X=0 \ .
\ee
This constraint  makes sense in Newtonian theory, but is not necessarily justified from a microphysics point of view, as it is baryon number, rather than mass that is conserved in a given nuclear reaction.

Finally, the total energy creation rate per unit volume is given by
\be
\epsilon^{\mathrm{ext}}=\sum_\X\left[v_\X^if_i^\X+D^{\X}_{ij} \nabla^j v_\X^i+\left(\mu_\X-\frac{1}{2}m_\X v_\X^2\right)\Gamma_\X\right]\ .
\ee

So far the model is quite general, but we now want to make direct contact with thermodynamics. In doing so, it makes sense to highlight the entropy component. We will, after all demand that the various dissipation channels adhere to the second law.
For an isolated system we have $\epsilon^{\mathrm{ext}}=0$, so that if we consider one of our fluids to represent the entropy of the system \footnote{Throughout the discussion we consider the regime where the thermal excitations of the system  can be treated as a fluid, i.e. we assume that the phonon mean free path is sufficiently short that this makes sense \citep{helium,heat,cfl}.} (we  denote this component by the subscript $\s$), the above relation leads to
\be
T\Gamma_\s=-f_iv_\s^i-D_{ij}\nabla^j v_s^i-\sum_{\X\neq\s}\left(\Gamma_\X\mu^\X+\hat{f}_i^\X w^i_{\X\s}+D^\X_{ij}\nabla^i w_{\X\s}^j\right)\ .\label{entropy}
\ee
where
\be
\hat{f}_i^\X=f_i^\X-{1\over 2} m^\X \Gamma_\X g_{ij}(v_\X^j+v_\s^j)\ , 
\ee
$f_i=\sum_\X f_i^\X$ is the total force acting on the system, and $D_{ij}=\sum_\X D^\X_{ij}$. Note that, as we are considering a closed system in the following, we  take $f_i=0$. Not also that, the individual $D^\X_{ij}$ do not have to be symmetric, even though the sum  $D_{ij}$ must be. Finally, the entropy component is taken to be massless, and its chemical potential is the temperature $\mu_\s=T$.

\section{The Onsager approach}

Let us now move on to consider the general form of the dissipation coefficients. To do this we continue to follow the analysis in \cite{monster}, and make use of the Onsager symmetry principle \citep{Onsager}. Our analysis will, however, differ from that in \cite{monster}, as in that work the authors neglected (erroneously) a number of terms involving $\nabla_i v_\s^j$. Hence, the resulting dissipative model was not as general as it should have been. This has already been noted in \cite{helium}, where  the analysis was reworked in the case of two fluids, one of which represented the entropy. As we intend to apply the discussion to the case of three (or more) fluids, we first consider the general form of the dissipation coefficients,  turning to particular examples later.

For any system, perturbations of the entropy density $s$ away from equilibrium must be given by quadratic deviations. This allows us to write \citep{monster}
\be
s\approx s_{\mathrm{eq}}-\frac{\Delta t}{2 T}\sum_{a,b} X_a L^{ab} X_b\ ,
\label{ds}\ee
which can be expressed,  in terms of the entropy creation rate $\Gamma_\s$, as
\be
T \Gamma_\s = -\frac{1}{2}\sum_{a,b} X_a L^{ab} X_b=\sum_{a=1}^{N} J^a X_a\ ,
\label{entropy2} \ee
where the $X_a$ are known as ``thermodynamic forces''. They represent a measure of the departure of the system from equilibrium, while the ``thermodynamic fluxes'' 
\be
J^a=-{1\over 2}\sum_b L^{ab}X_b\ ,
\ee
 arise in response. The Onsager symmetry principle states that microscopic reversibility implies $L^{ab}=L^{ba}$.

By comparing equation (\ref{entropy2}) to equation (\ref{entropy}) we can, by constructing the most general form for the tensor $L^{ab}$ in terms of the thermodynamical forces in the system, obtain the most general description of the dissipative terms in the Euler equations; the stress tensor $D_{ij}^\X$, the forces $\hat{f}^\X_i$ and the terms associated with the reaction rates. With this in mind, let us write down the most general form for the tensor $L^{ab}$.

From equation (\ref{entropy}) it might be tempting to take the thermodynamic forces to be $w_{\X\s}$, $\nabla_j w^i_{\X\s}$, $\nabla_j v^i_\s$ and $\mu_\X$, as in \cite{monster}. This is not quite appropriate, however, as we need the forces to vanish when thermodynamic equilibrium is reached. Hence, we should not work with the
chemical potentials, which obviously  do not  vanish in equilibrium.
This point comes to the fore when we consider problems with reactions, as in the case of bulk viscosity in a multi-fluid setting \cite{hyperon}.
We  need to replace the chemical potential with a more suitable ``force'', and the affinity \cite{prigogine} is the natural choice. In the context of neutron stars, this point has been made  by Carter and Chamel \cite{chamel}. The following analysis combines the key points of their discussion with the general multi-fluid formalism from the previous section.

Suppose there are $N$ total reactions among $M$ various constituents
$\X$ of our multi-fluid system, to be characterized in the usual way as
stoichiometric relations between the particle number densities~\footnote{Technically speaking one should consider mole numbers in these relations. However, for the kind of reactions that we consider in neutron star cores there is no difference.} $\nu^\X=n^\X/\left(\sum_\X n^\X\right)$ ; i.e.
\be
   \sum_{\X }^M {\rm R}_\X^I~\nu^\X \to \sum_{\X }^M {\rm P}_\X^I~\nu^\X
           \quad , \quad I = 1,...,N\ , 
\ee
where ${\rm R}_\X^I$ and ${\rm P}_\X^I$ are, respectively, the reactant
and product stoichiometric coefficients. The affinity $A^I$ of the
$I^{\rm th}$ reaction is then defined as
\be
    A^I \equiv \sum_{\X }^M \left({\rm R}_\X^I - {\rm P}_\X^I\right)
        {\mu^\X} \ .
\ee
At thermodynamic equilibrium the affinities vanish, which is why they make
appropriate thermodynamic forces.

 While it is clear that the affinities provide a natural description of the problem, it is important to recognize that the 
formulation is not quite complete at this point. In particular, the chemical potentials $\mu^\X$ 
become somewhat ambiguous in a multi-fluid context. In principle, the
chemical potentials should be defined as the energy per particle in the reference frame where the
chemical (or nuclear) reactions occur \cite{chamel2}. However, a multi-fluid mixture is characterized by
the presence of distinct velocity fields, neither of which provides the required frame. 
The relevant frame may, in fact, not be known a priori. The formulation we consider assumes an expansion away from ``equilibrium'', which ultimately involves
both dynamical and chemical considerations. The equilibrium frame may well depend on the dynamical evolution of the whole system. This complicates the issue considerably. Of course, in many situations of practical interest this problem may not be too severe. This is particularly the case when the relative velocities between the different fluid frames are small enough that it makes sense to linearise the problem. Noting that a satisfactory solution to the conceptual problem still remains to be developed, we proceed on the assumption that  a low-velocity model makes sense. 

According to Hess's Law, for each chemical reaction there is only one
thermodynamic variable to track in order to determine the changes; namely,
the ``degree of advancement'' $\xi_I$ for the various reactants. For each of
the $I = 1...N$ reactions, a variation $\Delta \xi_I$ corresponds to a
variation $\Delta \nu^\X_I$ of the participating fluids:
\be
   \frac{\Delta \nu_I^{\rm r}}{{\rm R}_{\rm r}^I} = ... =
   \frac{\Delta \nu_I^{\rm s}}{{\rm R}_{\rm s}^I} =
  - \frac{\Delta \nu_I^{\rm u}}{ {\rm P}_{\rm u}^I} = ... =
  - \frac{\Delta \nu_I^{\rm v}}{ {\rm P}_{\rm v}^I} = \Delta \xi_I \ ,
\ee
where ${\rm r},...,{\rm s}$ and ${\rm u},...,{\rm v}$ represent the $\X$-components for which the
${\rm R}_\X^I$ and ${\rm P}_\X^I$ are non-zero. The (irreversible) change
$\Delta s$ in the entropy due to these reactions is given by
\be
   \Delta s = \frac{1}{T} \sum_{I = 1}^N A^I \Delta \xi_I \ .
\ee
By comparing with equation (\ref{ds}), we see that the $\Delta \xi_I$ represent the appropriate thermodynamic
``fluxes''.

The variations $\Delta \nu^\X$ of the individual number densities, in some time
interval $\Delta t$, can also be determined by
\be
    \Delta \nu^\X =  \Gamma_\X \Delta t \ ,
\ee
where $\Gamma_\X$ is the particle number creation rate.

Each of the $N$ reactions has a corresponding change $\nu^\X_I$ that
contributes to $\Delta \nu^\X$, with the net result (as $\Delta t \to 0$)
\be
   \frac{d \nu^\X}{d t} = \sum_I \left({\rm R}_\X^I - {\rm P}_\X^I\right)
                          \frac{d \xi_I}{d t} \ .
\ee
Hence,
\be
 \Gamma_\X =  \sum_I \left({\rm R}_\X^I - {\rm P}_\X^I\right)
                    \frac{d \xi_I}{d t} \ .
\ee

If we define the reaction ``velocity'' $V^I\equiv \frac{d \xi_I}{d t}$ to be the thermodynamical flux,  then the change in entropy due to the reactions is
\be
   \Delta s =\sum_{\X \neq \s} \mu^\X \Gamma_\X=  \sum_{\X \neq \s} {\mu^\X} \left[\sum_I
                \left({\rm R}_\X^I - {\rm P}_\X^I\right)
                \frac{d \xi_I}{d t}\right]= \sum_{I} A^I V_I \ .
\ee

In the general framework the thermodynamic forces will then be $A^I$, $w^i_{\X \s}$, $\nabla_j v^i_\s$, $\nabla_j w^i_{\X\s}$, and the corresponding fluxes are $-V_I$,$-\hat{f}_i^\X$, $-D_{ij}$ and $-D^\X_{ij}$. Given this, we can
 follow the strategy of \cite{monster} to construct the fluxes out of the forces, limiting ourselves to the inclusion of quadratic terms. By making use of the Onsager symmetry principle, and noting that conservation of angular momentum~\footnote{Note that there is no similar constraint on the individual $D^\X_{ij} $ coeffients, despite this being a common assumption in the literature.} requires $D_{ij}$ to be symmetric (see equation (22) of \cite{monster}), we then arrive at;

\bear
-V^I &=& \sum_{\X,\Y\neq\s}\sum_{J} \left[L_\X^{I J} A_J +\tilde{L}^J_{ij} \nabla_i v_\s^j +\tilde{L}^{\Y J}_{ij} \nabla^{i}w_{\X\s}^j\right]\ , \\
-\hat{f}^\X_i &=& \sum_{\Y\neq\s}\left[ L^{\X\Y}_{ij} w^j_{\Y\s}+\tilde{L}^{\X\Y}_{ijk}\nabla^j w^k_{\Y\s}  \right] \ , \\
-D_{ij} &=&\sum_{\Y\neq\s}\left[\sum_J \left(\tilde{L}^J_{ij} A_J\right) +L_{ijkl}\nabla^kv^l_\s + \tilde{L}^\Y_{ijkl}\nabla^kw^l_{\Y\s}\right] \ ,   \\
-D_{ij}^\X &=& \sum_{\Y\neq\s}\left[\sum_J \left(\tilde{L}^{\X J}_{ij} A_J\right) +\tilde{L}^{\X\Y}_{ijk}w^k_{\Y\s}+ \tilde{L}^\X_{ijkl}\nabla^kv^l_\s + {L}^{\X\Y}_{ijkl}\nabla^kw^l_{\Y\s}\right] \ ,\label{fluxes}
\eear

 with

 \bear
&&L_\X^{I J} = \gamma_\X^{I J}=\gamma_\X^{J I}\quad \quad \tilde{L}^J_{ij}=\tau^J g_{ij}\quad\quad \tilde{L}^{\Y J}_{ij}=\tau^{\Y J} g_{ij}\ , \\
&&L^{\X\Y}_{ij}=2\mathcal{R}^{\X\Y}g_{ij}=2\mathcal{R}^{\Y\X}g_{ij}\quad\quad \tilde{L}^{\X\Y}_{ijk}=\mathcal{S}^{\X\Y}\epsilon_{ijk}=\mathcal{S}^{\Y\X}\epsilon_{ijk}\ ,  \label{cough}\\
&&L_{ijkl}=\zeta^\s g_{ij}g_{kl}+\eta^\s\left(g_{ik}g_{jl}+g_{il}g_{jk}-\frac{2}{3}g_{ij}g_{kl}\right)\ , \\
&&\tilde{L}^\Y_{ijkl}=\zeta^\Y g_{ij}g_{kl}+\eta^\Y\left(g_{ik}g_{jl}+g_{il}g_{jk}-\frac{2}{3}g_{ij}g_{kl}\right)\ , \\
&&{L}^{\X\Y}_{ijkl}=\zeta^{\X\Y} g_{ij}g_{kl}+\eta^{\X\Y}\left(g_{ik}g_{jl}+g_{il}g_{jk}-\frac{2}{3}g_{ij}g_{kl}\right)+\frac{1}{2}\sigma^{\X\Y}\epsilon_{ijm}\epsilon^m_{\ \ kl} \nonumber \\
&&\quad\quad\;\;=\zeta^{\Y\X} g_{ij}g_{kl}+\eta^{\Y\X}\left(g_{ik}g_{jl}+g_{il}g_{jk}-\frac{2}{3}g_{ij}g_{kl}\right)+\frac{1}{2}\sigma^{\Y\X}\epsilon_{ijm}\epsilon^m_{\ \ kl} \ . 
\label{olddiss}
 \eear

These relations suggest that the complete set of dissipation coefficients is given by
$$
\gamma_\X^{I J},\ \tau^J,\tau^{\Y J},\ \mathcal{R}^{\X\Y},\  \mathcal{S}^{\X\Y},\  \zeta^\s,\ \eta^\s,\ \zeta^\Y,\ \eta^\Y\, \zeta^{\X\Y},\ \eta^{\X\Y},\ \mbox{and}\ \sigma^{\X\Y} \ , 
$$
essentially the same as in \cite{monster}. The important difference is that, not only have we now correctly identified
the affinities as the thermodynamic forces, but we also have a host of new terms relating to the gradients of the entropy velocity $v_\s^i$.
These were neglected in \cite{monster}. An already complex problem has been made richer...

The problem may seem almost unmanageable at this point. In particular, how are we going the determine all the different dissipation coefficients? The answer is that we need to complement the phenomenological multi-fluid model with insights from microphysics (e.g. kinetic theory). Once we turn to that question we see that the complexity of the problem may reduce drastically. In fact, it is easy to argue that many of the different coefficients in the model will have the same microphysics origin (although various geometrical factors may differ). To see this, let us consider an example. Let us ask how particle scattering enters the problem. In a single-component system, the answer is simple. Scattering leads to friction that prevents fluid element from shearing, i.e. determine the coefficient of shear viscosity. The multi-fluid setting is more complicated, because we have to account for scattering both between particles of the same species and inter-species scattering. The former will (again) lead to the familiar shear viscosity from the Navier-Stokes equations, one term for each fluid species. The inter-species scattering affects the relative degrees of freedom. Two fluids can flow linearly through one another, and they can also have relative shear and expansion. The corresponding coefficients will all relate back to the scattering rates. A similar analysis relates to the various bulk viscosities. In this case, the problem reduces because the dissipation originates either from the relevant reaction rates (the case of main importance for neutron stars) or from fluctuations of the internal degrees of freedom for each species (as in water). A useful example, with direct relevance for one of the models discussed below, has been analysed by Gusakov and Kantor \cite{Ghyp} (although it should be noted that a translation between their model and that considered here is non-trivial). Other interesting discussions can be found in \cite{megus} and \cite{manna}.

One may also simplify the problem by constraining the physics. In the present context, the most relevant   constraint is associated with superfluidity, and we now turn to the corresponding problem.

\section{The superfluid constraint}

Let us now assume that one of the fluids (labelled ${\sf}$ in the following, not to be confused with the entropy component which is represented by a lowercase s) is superfluid. In this case we would expect it to be,
at least in the limit of low temperatures and velocities, irrotational.
Following \cite{Prix} and generalising the work of \cite{helium}, we impose the irrotational constraint on the momentum of this fluid. This means that we require
\be
\epsilon^{ijk} \nabla_j p_k^\sf=0\ ,
\ee
which leads to the constraint
\be
\nabla_i\Psi=\frac{1}{n_\sf}(\hat{f}_i^\sf-\nabla^j D^{\sf}_{j i})\ ,\label{sfcon}
\ee
for some scalar $\Psi$. In order to analyse this constraint it is useful to rewrite equation (\ref{entropy})
in terms of the variables $j_{\X \s}=n_\X w^i_{\X \s}$. This leads to
\be
T\Gamma_\s=-D_{ij}\nabla^i v^j_\s-\sum_{\X\neq\s} \left( \mathcal{F}^\X_i j_{\X\s}+\mathcal{D}^\X_{ij}\nabla^ij^j_{\X\s}\right) +\sum_I A_I V^I\ ,
\ee
where
\be
\mathcal{D}^\X_{ij}=\frac{1}{n_\X} D^\X_{ij} \quad\quad\mbox{and}\quad\quad \mathcal{F}_i^\X=\frac{1}{n_\X}\left[f^\X_i-\left(\frac{\nabla^j n_\X}{n_\X}\right)D^\X_{ji} \right]\ .
\ee
The constraint in (\ref{sfcon}) thus takes the form
\be
\nabla_i\Psi=\mathcal{F}_i^\sf-\nabla^j\mathcal{D}^\sf_{ij} \ .
\ee
Repeating the analysis of the thermodynamic fluxes (from  the previous section) in terms of the new variables, we find that
\bear
-V^I &=& \sum_{\X,\Y\neq\s} \sum_{J} M_\X^{I J} A_J +\tilde{M}^J_{ij} \nabla_i v_\s^j +\tilde{M}^{\Y J}_{ij} \nabla^{i}j_{\Y\s}^j\ , \\
-\mathcal{F}^\X_i &=& \sum_{\Y\neq\s} M^{\X\Y}_{ij} j^j_{\Y\s}+\tilde{M}^{\X\Y}_{ijk}\nabla^j j^k_{\Y\s}  \ , \\
-D_{ij} &=& \sum_{\Y\neq\s}\sum_J \tilde{M}^J_{ij} A_J +M_{ijkl}\nabla^kv^l_\s + \tilde{M}^\Y_{ijkl}\nabla^kj^l_{\Y\s} \ , \\
-\mathcal{D}^\X_{ij} &=& \sum_{\Y\neq\s}\sum_J \tilde{M}^{\X J}_{ij} A_J +\tilde{M}^{\X\Y}_{ijk}j^k_{\Y\s}+ \tilde{M}^\X_{ijkl}\nabla^kv^l_\s + {M}^{\X\Y}_{ijkl}\nabla^kj^l_{\Y\s} \ . \label{fluxes2}
\eear
(where the $M$ coefficients essentially replace the $L$'s from the previous section).
The constraint in (\ref{sfcon}) now implies that, for the superfluid component one must have
\be
\nabla_i\Psi=\nabla^j\left(\tilde{M}^{\sf J}_{ij} A_J+M^{\sf \Y}_{ijkl}\nabla^k j^l_{\Y\s}+\tilde{M}^\sf_{ijkl}\nabla^k v^l_\s\right)+\left(\nabla^j\tilde{M}^{\sf\Y}_{ijk}-M^{\sf\Y}_{ik}\right) j^k_{\Y\s}\ ,
\ee
which implies
\be
M^{\sf\Y}_{ik}=\nabla^j\tilde{M}^{\sf\Y}_{ijk}=0\ .
\ee
To see that these coefficients must vanish individually, consider Eq.~\eqref{cough}. 
The coefficient $M_{ik}^{\sf\Y}$  has the same form as $L_{ik}^{\X\Y}$ in Eq.~\eqref{cough}, i.e. it is symmetric, while the $\tilde{M}_{ijk}^{\sf\Y}$ is  antisymmetric when $i$ and $k$ are interchanged. 

The superfluid constraint thus takes the form
\be
\nabla_i\Psi=\nabla^j\left(\tilde{M}^{\sf J}_{ij} A_J+M^{\sf \Y}_{ijkl}\nabla^k j^l_{\Y\s}+\tilde{M}^\sf_{ijkl}\nabla^k v^l_\s\right),
\ee
from which we infer that
\beq
\tilde{M}^{\X J}_{ij}&=&\hat{\tau}^\sf g_{ij}\ ,\\
M^{\sf \Y}_{ijkl}&=&\hat{\zeta}^{\sf\Y}g_{ij}g_{kl}\ ,\\
\tilde{M}^\sf_{ijkl}&=&\hat{\zeta}^{\sf}g_{ij}g_{kl}\ .
\label{newdiss}
\eeq
If we compare these results to the general expressions in (\ref{olddiss}) we see that imposing the superfluid constraint on one, or more, of the fluids significantly reduces the number of dissipation coefficients. 

It is important to appreciate that this reduction comes about once we identify the appropriate thermodynamical fluxes in the system. This highlights the fact that there may be different ways of formulating any given problem, possibly leading to systems of seemingly different complexity. This is somewhat unfortunate, but we are not aware of any general prescription for avoiding this ambiguity. It is also worth noting that it is generally not meaningful to ``translate'' the coefficient in the two models we have provided. The basic reason for this is that the two systems are ``separated'' by a phase-transition (at the critical temperature for superfluidity). The models we have outlined apply either above, or well below, the relevant temperature. The increasing role of thermal excitations as the transition is approached make the corresponding problem tricky. In principle, the system must ``switch on'' the various dissipation channels that were removed by the superfluid constraint as the critical temperature is approached (reverting to the general system discussed in the previous section). 

The reduction associated with the superfluid problem may only be of formal interest, however. In practice, the irrotational constraint is too severe since a superfluid can rotate by forming an array of vortices. This \underline{complicates} the problem rather than simplifying it. First of all, we need to use the expressions in (\ref{olddiss}). Secondly, we need to worry about additional dissipation channels that come into play when vortices are present. Having said that, there may be some merit to an argument that one can take the irrotational model as starting point, adding only the particular mechanisms that are due to the vortices. In such a model, which may turn out to be accurate in many situations, the reduction of complexity due to the irrotational constraint is helpful. Most current discussions of neutron star vortex dynamics build on this idea.

The main problem with the irrotational assumption is that it means that we eliminate the vortex-mediated mutual friction, which arises from the balance between the Magnus force and the dissipative drag forces (such as electron scattering of vortex cores) on the vortices  \citep{mf2,Mendellb,trev}. This is known to be a key mechanism in neutron star dynamics.
We will not discuss the mutual friction in detail here, but it is nevertheless worth making a few general remarks. In the standard
two-fluid system of neutrons and protons, and if one considers straight vortices, the friction between a superfluid component $\X$ and another component $\Y$ can be taken into account by explicitly including a force of the form:
\be
f^{\X\Y}_i=\mathcal{B}^{'}\rho_\X n_{\mathrm{v}}\epsilon_{ijk}\kappa^j w_{\X\Y}^k+\mathcal{B}\rho_\X n_{\mathrm{v}}\epsilon_{ijk}\epsilon^{klm}\hat{\kappa}^j\kappa_l w_m^{\X\Y} \ , 
\label{mf}
\ee
where $\kappa^i$ is a vector aligned with the vortices, with magnitude $\kappa=h/2m_\X$~\footnote{Note that in general $\kappa=h/m_p$ where $m_p$ is the mass of the "boson" that forms the condensate. In the case of fermions that we are considering in this context $m_p$ is the mass of the Cooper pair, so that $m_p=2 m_\X$.} (a hat indicates a unit vector) and $n_\mathrm{v}$ is the vortex number density per unit surface area. $\mathcal{B}$ and $\mathcal{B}'$ are coefficients that encode the strength of the mechanism, but only the former is associated with actual dissipation. Given this additional force we see that, in the presence of vortices our analysis of the dissipation coefficients in the system is formally incomplete. We could make our analysis  more general by accounting for  the preferred direction associated with $\kappa^i$ when we design the dissipation terms. This extended model would obviously allow for the  standard mutual friction represented by (\ref{mf}), but also for the presence of a (significant) number of additional dissipative terms coupling the flows to the vorticity (see \citep{helium} for a discussion). 

Before moving on, it is also worth noting that 
the standard form for the mutual friction force may not be entirely appropriate.
In many circumstances the superfluid flow is expected to be turbulent. This means that the vortices are no longer ``straight'' but form a complicated tangle. In this case one can no longer use the expression in (\ref{mf}). Alternative forms in which the force is proportional to the cube of the relative velocity have been proposed. The form of this turbulent force and the coefficients involved are not well established in the neutron star context, although there have  been recent attempts to understand the relevance of the effect \citep{Mp1, Mp2,trev2}.

In fact, 
the nature of mutual friction in a general multi-fluid system may be considerably more complex. One may have to account for the interaction between several kinds of vortices which can form in the different superfluid or superconducting condensates. For example, \cite{babaev} has recently shown that for strong entrainment, or in the presence of superfluid $\Sigma^{-}$ hyperons, the usual picture in which one has rotation-induced neutron vortices, but not proton vortices, does not hold. Rather one can have ``composite''
vortices, strongly reducing the interaction between the superfluid and the magnetic-field carrying superconducting components of the star. Another 
possible complication has been discussed in \cite{zit}. 
Clearly, more work  is needed in order to understand how to include such concepts in our general picture.

\section{Perturbations}

The development of the dissipative multi-fluid formalism is obviously somewhat abstract, and we need to consider explicit examples in order to 
see how it can be applied. 
Ultimately, we are interested in how  dissipation affects the modes of oscillation of a rotating superfluid
neutron star. Given this, and the fact that dissipation is associated with deviations from equilibrium,
let us  consider linear perturbations of the multi-fluid equations of motion, (\ref{euler}) and (\ref{continuity}). Assuming that we are interested in rotational 
instabilities, like that of the Coriolis driven r-mode \cite{sfr1,rmode,hyperon}, we  work in the slow rotation
approximation to linear order, meaning that we perturb around a spherical background in which all fluids co-rotate
 and are in hydrostatic and chemical equilibrium.
These assumptions simplify the background equations considerably, as there is no dissipation and no
terms involving relative velocities~\footnote{The general problem, with a rotational lag between different components, is vastly more complicated.}. One simply has to consider the standard equations for hydrostatic equilibrium:
\be
\nabla_i p + \rho\nabla_i \Phi = 0\ , 
\ee
where $\Phi$ is the gravitational potential (as before) and where total density and pressure are given by
\bear
\rho&=&\sum_\X \rho_\X\ , \\
\nabla_i p &=& \sum_\X n_\X \nabla_i\mu_\X \ . 
\eear
The perturbed Euler equations in the rotating frame can then be written in the form (representing Eulerian variations by $\delta$)
\be
\partial_t\delta\pi_i^\X+n_\X\nabla_i\delta\mu_\X+\delta n_\X\nabla_i\mu_\X+2\rho_\X\epsilon_{ijk}\Omega^j\delta v_\X^k+\delta\rho_\X\nabla_i\Phi+\rho_\X\nabla_i\delta\Phi=\delta f_i-\nabla^j\delta D^{\X}_{ij}\ ,\label{eulerpert}
\ee
where $\Omega^i$ is the angular velocity of the star and $\rho_\X=m_\X n_\X$.
The perturbed momentum $\delta\pi_i^\X$ will not in general be parallel to the perturbed velocity of the $\X$ component;
due to the entrainment effect it  acquires components along the relative flows. We have
\be
\delta\pi_i^\X=g_{ij}\left(m_\X n_\X \delta v_\X^j+2\sum_\Y \alpha^{\X\Y} \delta w_{\Y\X}^j\right) \ , 
\ee
where the $\alpha^{\X\Y}$ are the entrainment coefficients defined in (\ref{ent}), the perturbations of which we do not need to consider since
we are considering a co-moving background. We also need the perturbed continuity equations, which take the form
\be
\partial_t\delta\rho_\X+\nabla_j  \left( \rho_\X\delta v_\X^j \right) =m^\X\Gamma_\X\ ,\label{continuitypert}
\ee
where we have assumed that the background is in chemical equilibrium. In other words, we  take the reaction rates $\Gamma_\X$ to arise at the linear perturbation level.
This obviously  makes sense since the reactions are triggered by deviations from chemical equilibrium.

The relations in (\ref{eulerpert}) and (\ref{continuitypert})  have the same structure for each fluid and represent the equations of motions for the $\X$ coupled degrees of freedom. In many situations, it can be an advantage to make use of this ``symmetry''. At the same time, it is
 instructive to show that the equations can be combined to regain the usual Navier-Stokes equations. To do this,
we sum the perturbed continuity equations in (\ref{continuitypert}) and assume mass conservation (i.e. $\sum_\X m^\X \Gamma_\X=0$).
This leads to
\be
\partial_t \delta\rho+\nabla_j(\rho\delta v^j)=0\ .\label{continuityall}
\ee
where we have introduced the velocity, $v^i$, associated with the total mass flux;
\be
\rho v^i =\sum_\X\rho_\X v_\X^i\ .
\ee
Meanwhile,
the sum of the Euler equations in (\ref{eulerpert}) leads to:
\be
\rho \partial_t  \delta v_i  +\nabla_i\delta p+2\rho \epsilon_{ijk}\Omega^j\delta v^k-\frac{\delta\rho}{\rho}\nabla_i p+\rho\nabla_i\delta\Phi=-\nabla^j \delta D_{ij}\ ,\label{eulerall2}
\ee
where $\delta D_{ij}=\sum_\X\delta D^\X_{ij}$.

Clearly, equations (\ref{continuityall}) and (\ref{eulerall2}) only account for one of the $N$ dynamical degrees of freedom, and one has to supplement them with the equations of motion for the remaining $N-1$ degrees of freedom, either directly from equations (\ref{eulerpert}) and (\ref{continuitypert}) or with suitable combinations of these.
In most previous work,  focused on the two-fluid case \citep{lm1,lm2,fmode,rmode}, it has been found advantageous to work with
the difference of the Euler equations. This leads to  an evolution equation for the relative velocity, sourced by the deviation from chemical equilibrium. In the general case, where one has more fluids and different kinds of reactions it is not so obvious what the best combination to use may be. One may have to consider the issue on a case by case basis.

Before we consider explicit examples, it is worth commenting on one particular problem area where the present results
may be applied. Problems in neutron star dynamics are closely linked to the effort to detect gravitational waves.
A key problem concerns oscillations and instabilities of rotating compact stars. In order to consider neutron star models with realistic interior
composition, one needs a formalism able to determine the timescale on which oscillations of superfluid neutron stars are damped out by various dissipative processes.
 There are essentially two approaches to this problem. The first consists of solving the full dissipative problem, essentially ``integrating'' equations (\ref{eulerpert}) and (\ref{continuitypert}). Given the complexity of the different dissipative terms this approach is, however, often not viable.
The second option is to estimate the various dissipation timescales from energy integrals, see \cite{fmode} for discussion.
This approach is based on the assumption, valid in many physical circumstances, that the dissipation is weak and does not significantly alter the nature of the
solutions to the conservative problem (in which the dissipative terms are absent). This will  be the case when the damping timescale is significantly longer than the dynamical timescales we are interested in, i.e. the oscillation periods considered.
In this case one ``simply'' has to solve the continuity equations in (\ref{continuitypert}) and Euler equations in (\ref{eulerpert}) without the dissipative terms $D^\X_{ij}$ and force terms $f^{\X}_i $. In addition, one requires an energy for the system.
The relevant object can be obtained by multiplying the Euler equations for each component  by $\rho_\X v_\X^{i*}$ (where the $^*$ represents complex conjugation) and adding it to its complex conjugate. Combining the contributions leads to a total time derivative of a quantity which we can define as the energy, $E$.
The time derivative of this energy $\partial_t E$ follows from the ``right-hand side'' of the dissipative equations of motion, and allows us to estimate the
damping timescale
\be
\tau \approx {2E \over  \partial_t E}  \ .
\ee
Examples of this kind of analysis are discussed in \cite{NilsRev}.

\section{Examples: Exotic neutron star cores}

So far, our discussion has mainly concerned the general multi-fluid formalism. Once we move beyond the single-fluid model
the situation clearly becomes very complex. The main lesson is that we need to consider (at least in principle) a plethora of
new dissipation channels. In order to gain better insight into this reality, we will consider two
problems with direct relevance for neutron star astrophysics. Both problems concern the deep neutron star core. In the first case we assume that the core has a sizeable hyperon component
while the second example concerns a deconfined quark core. The examples are similar in that they both require us to consider three
coupled ``fluids''. Yet, they are sufficiently different to illustrate the subtleties of these kinds of problems.

\subsection{Hyperon cores}

The first of our examples concerns a neutron star with a hyperon core. Several proposed equations of state predict the appearance of hyperons at supranuclear densities (see e.g. \cite{Glenn,LO}). The recent measurement of a neutron star mass of 1.97 $M_\odot$ \cite{demorest}  appears
to place stringent constraints on  equations of state with softening components, like hyperons, but it is important to keep in mind that the presence of hyperons is  expected for fundamental physics reasons \cite{stone} and
the models remain incomplete until the many-body interactions are fully accounted for. As this is an immensely difficult problem, it is relevant to consider indirect evidence for the presence (or, indeed, absence) of hyperons in the core of astrophysical neutron stars.  The r-mode instability may provide interesting constrains in this respect.

The $\Lambda$ and $\Sigma^{-}$ hyperons are predicted to have the lowest thresholds for formation. The resultant problem
is of great interest, in particular for gravitational-wave physics, as the presence of hyperons increases the strength of the bulk viscosity and
reduces the part of parameter space in which gravitational-wave driven instabilities may operate. The
exact details of the damping can  have  observational consequences. Two of us have recently examined the effect of the additional damping coefficients
on the r-mode instability \cite{hyperon}. That analysis was based on the simpler case of a $\Sigma^{-}$ hyperon core, which can be cast as a two-fluid problem.
In the following we develop present the formalism for a core in which both $\Lambda$ and $\Sigma^{-}$ hyperons are present.
As we will see, this is essentially a three-fluid problem.

We start from the equations of motion for a cold \underline{four} fluid system, formed by neutrons ($\n$), protons ($\p$), $\Sigma^{-}$ and $\Lambda$ (in the interest of clarity, we do not account for the presence of muons, even though they would be easily included in the model).
This means that we have already assumed that the electrons are locked to the protons on a much shorter timescale than the dynamical timescales
we are considering~\footnote{This is reasonable approximation as long as the characteristic frequency of the phenomena we consider
is below the electron-proton plasma and cyclotron frequency, which is the case for most problems of astrophysical interest.}, and neglect their mass compared to that of the other components \citep{Mendella,Mendellb}. For similar reasons,
the $\Sigma^{-}$ hyperons can be considered locked to the charged component. Hence, we let $v^i_\h=v^i_\p$ such that all remaining ``fluids'' are overall charge neutral (the scale considered is assumed to be larger than the relevant screening length).
We will, however, retain a separate $\Sigma^{-}$ mass fraction in the continuity equations. We do this in order to make the analysis of the bulk viscosity damping 
simpler. After all, the expectation \cite{LO,hyper} is that  the bulk viscosity will be dominated by hyperon creating processes like
\bear
\n+\n &\rightleftharpoons& \p + \Sigma^{-}\ ,\\
\n+\p&\rightleftharpoons& \p + \Lambda\ .
\label{react}
\eear

Moving on to the  perturbations of the hyperon core, we write the momenta of our three components ($\n$=neutrons and $\c$=protons ($\p$) locked to electrons and $\Sigma^{-}$ hyperons ($\h$), $\l=\Lambda$ hyperons) as
\bear
\pi^\n_{i}&=&g_{ij}\left[m_\n n_\n v_\n^j-2(\alpha^{\n\p}w_{\n\p}^j+\alpha^{\n\h}w_{\n\h}^j+\alpha^{\n\l}w_{\n\l}^j)\right],\\
\pi^\c_{i}&=&g_{ij}\left\{(m_\p n_\p+m_\h n_\h) v_\p^j+2[(\alpha^{\n\p}+\alpha^{\p\l})w_{\n\p}^j-\alpha^{\p\l}w_{\n\l}^j]\right\},\\
\pi^\l_{i}&=&g_{ij}\left\{m_\l n_\l v_\l^j+2[(\alpha^{\h\l}+\alpha^{\p\l}+\alpha^{\n\l})w_{\n\l}^j-(\alpha^{\h\l}+\alpha^{\p\l})w_{\n\p}^j]\right\}.\label{mom}
\eear
We can then write the perturbed Euler equations as one equation for the centre-of-mass velocity, equation (\ref{eulerall2}), and two equations for relative
velocities. As there will be no perturbations of the entrainment for co-rotating backgrounds, the two difference equations take the form:
\begin{multline}
\left(1-\bar{\varepsilon}_1-\frac{\bar{\varepsilon}_2\bar{\varepsilon}_4}{1-\bar{\varepsilon}_3}\right)\partial_t \delta w^{\n\p}_i+\nabla_i \left(\delta\tilde{\beta}_\c+\frac{\bar{\varepsilon}_2}{1-\bar{\varepsilon}_3}\delta\tilde{\beta}_\l\right)+2\epsilon_{ijk}\Omega^j \left(\delta w_{\n\p}^k+\frac{\bar{\varepsilon}_2}{1-\bar{\varepsilon}_3}\delta w_{\n\l}^k\right)=\\
=-\left(1+\frac{\bar{\varepsilon}_2}{1-\bar{\varepsilon}_3}-\frac{\rho_\n}{\rho_\h+\rho_\p}\right)\nabla_j\delta\tilde{\mathcal{D}}^{j\n}_i+\left(\frac{\bar{\varepsilon}_2}{1-\bar{\varepsilon}_3}-\frac{\rho_\l}{\rho_\h+\rho_\p}\right)\nabla_j\delta\tilde{\mathcal{D}}^{j\l}_i+\frac{1}{\rho_\h+\rho_\p}\nabla_j\delta{D}^j_i+\\
+\left(1+\frac{\bar{\varepsilon}_2}{1-\bar{\varepsilon}_3}+\frac{\rho_\n}{\rho_\h+\rho_\p}\right)\delta\tilde{\mathcal{F}}^\n_i-\left(\frac{\bar{\varepsilon}_2}{1-\bar{\varepsilon}_3}-\frac{\rho_\l}{\rho_\h+\rho_\p}\right)\delta\tilde{\mathcal{F}}^\l_i \ , 
\end{multline}

\begin{multline}
\left(1-\bar{\varepsilon}_3-\frac{\bar{\varepsilon}_2\bar{\varepsilon}_4}{1-\bar{\varepsilon}_1}\right)\partial_t\delta w^{\n\l}_i+\nabla_i \left(\delta\tilde{\beta}_\l+\frac{\bar{\varepsilon}_4}{1-\bar{\varepsilon}_1}\delta\tilde{\beta}_\c\right)+2\epsilon_{ijk}\Omega^j \left(\delta w_{\n\l}^k+\frac{\bar{\varepsilon}_4}{1-\bar{\varepsilon}_1}\delta w_{\n\p}^k\right)=\\
=-\left(1+\frac{\bar{\varepsilon}_4}{1-\bar{\varepsilon}_1}\frac{\rho_\n+\rho_\h+\rho_\p}{\rho_\h+\rho_\p}\right)\nabla_j\delta\tilde{\mathcal{D}}^{j\n}_i+\left(1-\frac{\bar{\varepsilon}_4}{1-\bar{\varepsilon}_1}\frac{\rho_\l}{\rho_\h+\rho_\p}\right)\nabla_j\delta\tilde{\mathcal{D}}^{j\l}_i\\
+\frac{1}{\rho_\h+\rho_\p}\frac{\bar{\varepsilon}_4}{1-\bar{\varepsilon}_1}\nabla_j\delta{D}^j_i-\\
-\left(1-\frac{\bar{\varepsilon}_4}{1-\bar{\varepsilon}_1}\frac{\rho_\l}{\rho_\h+\rho_\p}\right)\delta\tilde{\mathcal{F}}^\l_i+\left(1+\frac{\bar{\varepsilon}_4}{1-\bar{\varepsilon}_1}\frac{\rho_\n+\rho_\h+\rho_\p}{\rho_\h+\rho_\p}\right)\delta\tilde{\mathcal{F}}^\n_i\ .
\end{multline}
In these expressions the tildes ($\tilde\ $) indicate that the variable has been rescaled with the relevant mass $m^\X$, e.g. $\delta\tilde{\beta}_\c = \delta \beta_\c/m^\c$.
{ We  have accounted for the different masses of the individual components}, but it is worth noting that neglecting the mass difference of the hyperons, taking all masses to be equal, is a reasonable approximation if we want to determine the damping timescale of the r-mode instability \cite{hyperon}. This obviously
also makes the problem more tractable.

We have also defined various combinations of the entrainment parameters
\bear
&&\bar{\varepsilon}_1=\frac{2(\alpha^{\n\p}+\alpha^{\n\h})}{\rho_\n}+\frac{2(\alpha^{\n\p}+\alpha^{\p\l})}{\rho_\h+\rho_\p}\ ,\;\;\;\;\;\;\;\bar{\varepsilon}_2=\frac{2\alpha^{\n\l}}{\rho_\n}-\frac{\alpha^{\p\l}}{\rho_\h+\rho_\p}\ , \\
&&\bar{\varepsilon}_3=\frac{2(\alpha^{\n\l})}{\rho_\n}+\frac{2(\alpha^{\p\l}+\alpha^{\n\l}+\alpha^{\h\l})}{\rho_\l}\ ,\;\;\;\;\;\;\;\bar{\varepsilon}_4=\frac{2(\alpha^{\n\p}+\alpha^{\n\h})}{\rho_\n}-\frac{2(\alpha^{\h\l}+\alpha^{\p\l})}{\rho_\l}\ .
\eear
It is worth noting that, while the problem would simplify significantly if we were to ignore these coefficients we have physics justification for doing so. In fact, based on the familiar case of neutron-proton entrainment, there is every reason to expect the mechanism to be relevant also for the hyperons. This is, indeed, borne out by the results in \cite{Gusakov1,Gusakov2}.

From the perturbed continuity equations we derive the usual relation for the total density perturbations, equation (\ref{continuityall}), and three equations for the mass fractions
\bear
\partial_t\delta x_\p&=&-\frac{1}{\rho}\nabla_j\left\{ \rho x_\p[x_\l \delta w_{\n\l}^j-(1-x_\p-x_\h)\delta w_{\n\p}^j]\right\}-\delta v^j\nabla_j x_\p+\frac{m_\p\Gamma_\p}{\rho}\ , \\
\partial_t\delta x_\l&=&-\frac{1}{\rho}\nabla_j \left\{ \rho x_\l[(x_\p+x_\h) \delta w_{\n\p}^j-(1-x_\l)\delta w_{\n\l}^j]\right\}-\delta v^j\nabla_j x_\l+\frac{m_\l\Gamma_\l}{\rho}\ , \\
\partial_t\delta x_\h&=&-\frac{1}{\rho}\nabla_j\left\{ \rho x_\h[x_\l \delta w_{\n\l}^j-(1-x_\p-x_\h)\delta w_{\n\p}^j]\right\}-\delta v^j\nabla_j x_\h+\frac{m_\h\Gamma_\h}{\rho}\ ,
\eear
where we have introduced the fractions $x_\X = \rho_\X/\rho$.
Note that the above equations are not independent (otherwise one would have four, not three, fluids).  We have considered the hyperon and proton fractions separately in order to have the hyperon and proton creation rates explicit in the equations. The intention is to make the analysis of bulk viscosity simpler when one has to deal with the rates of several reactions, as in (\ref{react}). One then needs to apply the additional condition of charge neutrality, which implies that $n_\p=n_\h+n_\e$ where $n_\e$  is the number density of electrons. However, when the fraction of $\Sigma^-$ is significant the electron   population is depleted and we may be able to  use the approximate relation $\n_\p\approx n_\h$ \citep{hyperon}.

Let us now examine the nature of the dissipative terms in the  equations of motion. We shall assume that the neutrons and $\Lambda$ hyperons are superfluid, and thus apply the irrotational constraint (\ref{sfcon}) to both these components. Meanwhile, following \cite{monster}, we shall neglect heat conduction and assume that the entropy flows with the charged components, which thus represents the ``normal'' component. Note that we are explicitly taking $v_\h=v_\p$, so that $j_{\h\s}=j_{\p\s}=0$.
In this approximation the required thermodynamic fluxes take the form
\bear
-V^I &=& \sum_{\X} \sum_{J} M_\X^{I J} A_J +\tilde{M}^J_{ij} \nabla_i v_\p^j +\tilde{M}^{\n J}_{ij} \nabla^{i}j_{\n\s}^j+\tilde{M}^{\l J}_{ij} \nabla^{i}j_{\l\s}^j\ ,\\
-\mathcal{F}^\n_i &=& -\mathcal{F}^\l=0\ , \\
-D_{ij} &=& \sum_J \tilde{M}^J_{ij} A_J +M_{ijkl}\nabla^kv^l_\p+ \tilde{M}^\n_{ijkl}\nabla^kj^l_{\n\p}+ \tilde{M}^\l_{ijkl}\nabla^kj^l_{\l\p}\ ,  \\
-\mathcal{D}^\n_{ij} &=& \sum_J \tilde{M}^{\n J}_{ij} A_J + \tilde{M}^\n_{ijkl}\nabla^kv^l_\p + {M}^{\n\n}_{ijkl}\nabla^kj^l_{\n\p} + {M}^{\n\l}_{ijkl}\nabla^kj^l_{\l\p}\ , \\
-\mathcal{D}^\l_{ij} &=& \sum_J \tilde{M}^{\l J}_{ij} A_J + \tilde{M}^\l_{ijkl}\nabla^kv^l_\p + {M}^{\l\n}_{ijkl}\nabla^kj^l_{\n\p} + {M}^{\l\l}_{ijkl}\nabla^kj^l_{\l\p}\ ,\label{fluxes3}
\eear
with coefficients
\bear
&&M_\X^{IJ}=\gamma_\X^{IJ}\quad\quad\tilde{M}^J_{ij}=\tau^J g_{ij}\quad\quad\tilde{M}^{\n J}_{ij}=\tau^{\n J} g_{ij}\quad\quad\tilde{M}^{\l J}_{ij}=\tau^{\l J} g_{ij}\ , \\
&&M_{ijkl}=\zeta^\s g_{ij}g_{kl}+\eta^\s\left(g_{ik}g_{jl}+g_{il}g_{jk}-\frac{2}{3}g_{ij}g_{kl}\right)\ , \\
&&\tilde{M}^{\n}_{ijkl}=\hat{\zeta}^{\n}g_{ij}g_{kl}\quad\quad\tilde{M}^{\l}_{ijkl}=\hat{\zeta}^{\l}g_{ij}g_{kl}\ , \\
&&M^{\n\n}=\hat{\zeta}^{\n\n}g_{ij}g_{kl}\quad\quad M^{\n\l}=\hat{\zeta}^{\n\l}g_{ij}g_{kl}\quad\quad M^{\l\l}=\hat{\zeta}^{\l\l}g_{ij}g_{kl} \ .
\eear
We thus have the standard shear viscosity coefficient $\eta^\s$ that appears in the Navier-Stokes equations, but there are now 6 bulk viscosity coefficients and the reaction rates depend on the flows via the $\tau^J$ and $\tau^{\X J}$ coefficients.  
%Dissipative effects related to so-called ``rocket'' terms are also encoded in these parameters \cite{rocket}.
It is important to note that the new viscosity coefficients in the equations of motion are related to the relative flow. A generic perturbation of the system,
or indeed a typical oscillation mode, will be represented by the coupled degrees of freedom. It is by no means obvious from the outset to what extent the
various degrees of freedom are excited in a given situation.
A useful illustration is provided by the (polar) f-mode of a superfluid neutron stars, for which the damping is dominated by the vortex mutual friction and hence
relies on the relative motion of the two components \cite{fmode}. The damping can be strong because the excitation of the relative motion is considerable.
This is in clear contrast to the results for the (axial) r-modes \cite{rmode}, for which the exitation of the relative flow is weak (at least for the mode that is the
strongest gravitational-wave emitter).

\subsection{CFL and kaons}

Our second  example concerns a core of deconfined quarks in the colour-flavour locked (CFL) phase combined with a population of kaons. This problem is of great interest as the ground state of matter in neutron star cores has been the object of  vigorous investigation (see e.g. \cite{reviewA} for a review). While the pure CFL phase (which can be conveniently be described by a two-fluid model with a CFL condensate coupled to phonons \cite{cfl}) is the ground state of cold matter at asymptotically high densities, it is thought that a kaon condensate is likely to be present
at realistic neutron star core densities (leading to the so-called CFL-K$^0$ phase \citep{alf1}). This possibility has recently been considered in connection with the damping of the r-mode instability \citep{RJ}, but the multi-fluid aspects of the problem have (so far) been ignored.

In order to model this situation we  need to consider a three-fluid system given by the (neutral) CFL condensate, the kaons and the excitations of the system, which we treat as a massless entropy fluid (as in \cite{heat,cfl}). The analysis then proceeds (essentially) as in the previous section. We indicate the CFL condensate with $\c$, the kaons with $\k$ and the entropy with $\s$. The momenta of the different fluids are
\bear
\pi^\c_{i}&=&g_{ij}\left[m_\c n_\c v_\c^j-2(\alpha^{\c\k}w_{\c\k}^j+\alpha^{\c\s}w_{\c\s}^j)\right]\ , \\
\pi^\k_{i}&=&g_{ij}\left\{m_\k n_\k v_\k^j+2[(\alpha^{\c\k}+\alpha^{\k\s})w_{\c\k}^j-\alpha^{\c\s}w_{\c\s}^j]\right\}\ ,\\
\pi^\s_{i}&=&2 g_{ij} \left[(\alpha^{\c\s}+\alpha^{\k\s})w_{\c\s}^j-\alpha^{\k\s}w_{\c\k}^j\right]\ ,\label{moms}
\eear
Note that, as we impose that the fluids flow together in the background,  the unperturbed entropy momentum (\ref{moms}) vanishes.
From the Euler equations (\ref{euler}) one then finds that
\be
s\nabla_i T=0 \ ,
\ee
which implies that the unperturbed core has uniform temperature (we have made the identification $n_\s=\s$, $\mu_\s=T$). The background system is  isothermal.

We can again write the perturbed Euler equations as one equation for the total velocity, equation (\ref{eulerall2}), and two equations for the counter-moving velocities:
\begin{multline}
\partial_t\delta w^i_{\c\s}=\left[\frac{\varepsilon_T(\varepsilon^{\k}-\varepsilon^{\s\c})}{\tilde{\epsilon}}-1\right]\frac{s}{\bar{\alpha}}\nabla^i\delta T-\frac{\varepsilon_\mathrm{T}}{\tilde{\epsilon}}(\nabla^i\delta\tilde{\beta}+2\epsilon^{ijk}\Omega_j \delta w^{\c\k}_k)-\frac{1-\varepsilon_{\c\k}-\bar{\varepsilon}}{\tilde{\varepsilon}\bar{\alpha}}\nabla_j \delta D^{ij}\\
+\left(\frac{\rho_\c}{\bar{\alpha}}\frac{1-\varepsilon_{\c\k}-\bar{\varepsilon}}{\tilde{\varepsilon}}-\frac{\varepsilon_\mathrm{T}}{\tilde{\varepsilon}}\right)\nabla_j\delta\tilde{\mathcal{D}}^{ij}_{\c}+\left(\frac{\rho_\k}{\bar{\alpha}}\frac{1-\varepsilon_{\c\k}-\bar{\varepsilon}}{\tilde{\varepsilon}}+\frac{\varepsilon_\mathrm{T}}{\tilde{\varepsilon}}\right)\nabla_j\delta\tilde{\mathcal{D}}^{ij}_{\k}\\
+\left(\frac{\varepsilon_{\mathrm{T}}}{\tilde{\varepsilon}}-\frac{\rho_\c(1-\varepsilon_\k-\bar{\varepsilon})}{\bar{\alpha}\tilde{\varepsilon}}\right)\delta\tilde{\mathcal{F}}_\c^i-\left(\frac{\varepsilon_{\mathrm{T}}}{\tilde{\varepsilon}}+\frac{\rho_\k(1-\varepsilon_\k-\bar{\varepsilon})}{\bar{\alpha}\tilde{\varepsilon}}\right)\delta\tilde{\mathcal{F}}_\k^i\ ,
\end{multline}

\begin{multline}
\partial_t\delta w^i_{\c\k}=\left[\frac{(\varepsilon^{\k}-\varepsilon^{\s\c})}{\tilde{\epsilon}}\right]\bar{s}\nabla^i\delta T-\frac{1}{\tilde{\epsilon}}(\nabla^i\delta{\tilde\beta}+2\epsilon^{ijk}\Omega_j\delta w^{\c\k}_k)+\frac{\varepsilon_{\k}-\varepsilon_{\c\s}}{\tilde{\varepsilon}\bar{\alpha}}\nabla_j \delta D^{ij}\\
-\frac{\bar{\alpha}+\rho_\c(\varepsilon_\k-\varepsilon_{\c\s})}{\bar{\alpha}\tilde{\varepsilon}}\nabla_j\delta\tilde{\mathcal{D}}^{ij}_{\c}+\frac{\bar{\alpha}-\rho_\k(\varepsilon_\k-\varepsilon_{\c\s})}{\bar{\alpha}\tilde{\varepsilon}}\nabla_j\delta\tilde{\mathcal{D}}^{ij}_{\k}\\
+\frac{\bar{\alpha}+\rho_\c(\varepsilon_\k-\varepsilon_{\c\s})}{\bar{\alpha}\tilde{\varepsilon}}\delta\tilde{\mathcal{F}}^{i}_{\c}-\frac{\bar{\alpha}-\rho_\k(\varepsilon_\k-\varepsilon_{\c\s})}{\bar{\alpha}\tilde{\varepsilon}}\delta\tilde{\mathcal{F}}^{i}_{\k} \ ,
\end{multline}

where ${\delta\tilde{\beta}}=\delta\mu_\c/m_\c-\delta\mu_\k/m_\k$ 
\bear
&&\varepsilon^{\c\k}=\frac{2\alpha^{\c\k}}{\rho_\c}\quad\quad\varepsilon^{\c\s}=\frac{2\alpha^{\c\s}}{\rho_\c}\quad\quad\bar{\varepsilon}=\frac{2(\alpha^{\c\k}+\alpha^{\k\s})}{\rho_\k}\quad\quad\bar{\alpha}={2(\alpha^{\c\s}+\alpha^{\k\s})}\ , \\
&&\varepsilon^{\k}=\frac{2\alpha^{\c\s}}{\rho_\k}\quad\quad\varepsilon_\mathrm{T}=\frac{\alpha^{\k\s}}{\alpha^{\c\s}+\alpha^{\k\s}}\quad\quad\tilde{\varepsilon}=1-\varepsilon^{\c\k}-\bar{\varepsilon}+\varepsilon_\mathrm{T}(\varepsilon^\k-\varepsilon^{\c\s})\ .
\eear
Since the background is co-moving,  perturbations of the  entrainment parameters do not appear in the equations of motion.
In contrast is the previous example, most of the required coefficients have not been previously considered in the literature. Hence, we do not know if it necessary to retain the
entrainment between quarks and kaons. The entropy entrainment, on the other hand, is known to be key to the thermal relaxation in the problem \cite{heat}. The effect may be weak at low temperatures, but there are fundamental arguments for why it should be present.

From the perturbed continuity equations we derive the usual relation for the total density perturbations, equation (\ref{continuityall}), a continuity equation for the
kaon mass fraction ($x_\k=\rho_\k/\rho$)
\be
\partial_t\delta x_\k=\frac{1}{\rho}\nabla_i\left[x_\k(1-x_\k)\delta w_{\c\k}^i\right]-\delta v^i\nabla_i x_\k+\frac{m_\k\Gamma_\k}{\rho}\ ,
\ee
and a conservation law for entropy
\be
\partial_t\delta s+\nabla_i (s \delta v_\s^i)=\partial_t\delta s+\nabla_i s(\delta v^i+x_\k \delta w_{\c\k}^i-\delta w_{\c\s}^i)=0\ .
\ee
Here it is worth noting that, since we are working at the linear perturbation level and the background is isothermal (i.e. there is no heat flow
in the unperturbed configuration) the problem is adiabatic. 

Let us now turn to the dissipative terms in the Euler equations. If we assume that the CFL condensate and the kaons are  superfluid we can apply the irrotational constraint (\ref{sfcon}) to both these components. Meanwhile, the entropy represents the ``normal'' fluid. The situation is thus very similar to that for hyperons, and the fluxes are explicitly given by;
\bear
-V^I &=& \sum_{\X} \sum_{J} M_\X^{I J} A_J +\tilde{M}^J_{ij} \nabla_i v_\s^j +\tilde{M}^{\c J}_{ij} \nabla^{i}j_{\c\s}^j+\tilde{M}^{\k J}_{ij} \nabla^{i}j_{\k\s}^j\ ,\\
-\mathcal{F}^\c_i &=& -\mathcal{F}^\k=0\ , \\
-D_{ij} &=& \sum_J \tilde{M}^J_{ij} A_J +M_{ijkl}\nabla^kv^l_\s+ \tilde{M}^\c_{ijkl}\nabla^kj^l_{\c\s}+ \tilde{M}^\k_{ijkl}\nabla^kj^l_{\k\s}\ , \\
-\mathcal{D}^\c_{ij} &=& \sum_J \tilde{M}^{\c J}_{ij} A_J + \tilde{M}^\c_{ijkl}\nabla^kv^l_\s + {M}^{\c\c}_{ijkl}\nabla^kj^l_{\c\s} + {M}^{\c\k}_{ijkl}\nabla^kj^l_{\k\s}\ , \\
-\mathcal{D}^\k_{ij} &=& \sum_J \tilde{M}^{\k J}_{ij} A_J + \tilde{M}^\k_{ijkl}\nabla^kv^l_\s + {M}^{\k\c}_{ijkl}\nabla^kj^l_{\c\s} + {M}^{\k\k}_{ijkl}\nabla^kj^l_{\k\s}\ ,\label{fluxes3a}
\eear
with
\bear
&&M_\X^{IJ}=\gamma_\X^{IJ}\quad\quad\tilde{M}^J_{ij}=\tau^J g_{ij}\quad\quad\tilde{M}^{\c J}_{ij}=\tau^{\c J} g_{ij}\quad\quad\tilde{M}^{\k J}_{ij}=\tau^{\k J} g_{ij}\ , \\
&&M_{ijkl}=\zeta^\s g_{ij}g_{kl}+\eta^\s\left(g_{ik}g_{jl}+g_{il}g_{jk}-\frac{2}{3}g_{ij}g_{kl}\right)\ ,\\
&&\tilde{M}^{\c}_{ijkl}=\hat{\zeta}^{\c}g_{ij}g_{kl}\quad\quad\tilde{M}^{\k}_{ijkl}=\hat{\zeta}^{\k}g_{ij}g_{kl}\ ,\\
&&M^{\c\c}=\hat{\zeta}^{\c\c}g_{ij}g_{kl}\quad\quad M^{\c\k}=\hat{\zeta}^{\c\k}g_{ij}g_{kl}\quad\quad M^{\k\k}=\hat{\zeta}^{\k\k}g_{ij}g_{kl}\ .
\eear
Once again we have the standard shear viscosity coefficient  $\eta^\s$ and 6 bulk viscosity coefficients.

In the above analysis we accounted for a charge-neutral superfluid $K^0$ condensate. Another possibility is that, for sufficiently  hot neutron stars, there exists a sizable thermal population of kaons \citep{kaon1,kaon2}. This would change the  problem considerably, as one would expect that (at least as a first approximation) the thermal kaons would be locked to the entropy (the thermal excitations), thus adding to the ``normal'' fluid. Moreover, the thermal kaons would be massive. To deal with this situation, we could  once again solve a two-fluid problem, similar to that of a CFL condensate coupled to phonons \cite{cfl}. The main difference would be that the superfluid condensate would no longer be coupled to a massless fluid, but rather to the fluid formed by  thermal kaons and  phonons. The mass density of such a fluid would  be temperature dependent. This situation has not been considered  previously. It could be interesting, both from a conceptual and a practical point-of-view. There may, for example, be
  interesting consequences for the r-mode instability, as the Coriolis force which drives the mode will not act on a massless fluid (such as a phonon gas) but it would act on the thermal kaons. 

A similar problem that we need to consider is that of superfluid neutrons and protons coupled to thermal excitations \cite{thermal,gus}. In the simplest approximation one may assume that the system is well below the superfluid transition temperature in which case thermal effects can safely be neglected. This is, indeed, the assumption in most existing work on neutron star dynamics. However, given the strong density dependence of the superfluid pairing gaps there will \underline{always} be regions of the star that are close to the superfluid transition temperature. In these regions thermal effects are important. To model such regions one should consider a three-fluid system of superfluid neutrons, superconducting protons (locked to the electrons) and thermal excitations (the entropy). Formally, such a system would be very close to the CFL-K$^0$ core. It may be sufficient to replace the density of the CFL condensate with the neutron density and the kaon density with the proton density in all the above equations.

\section{Conclusions}

We have presented a general flux-conservative formalism for modelling dissipation in multi-fluid systems, extending and correcting the model from \cite{monster}. The formalism was developed with superfluid neutron star cores in mind, but is
sufficiently general that it can be applied to a variety of analogous multi-fluid systems.
The introduction of extra degrees of freedom (the separate ``fluids'') leads to a number of additional dissipation coefficients compared to standard single-fluid hydrodynamics. This may affect the dynamics of the system 
significantly, e.g. impact on the nature of the modes of oscillation of a neutron star. This may, in turn, have  repercussions for the gravitational waves emitted by the system.
In order to understand this effect we developed the most general form for the dissipation coefficients and discussed how their number can be reduced  by imposing the ``superfluid'' constraint on one or more fluids (imposing that the flow must remain irrotational).
This constraint is often too severe, as a superfluid condensate will not be macroscopically irrotational, but will rather mimic the effect of bulk rotation by creating an array of vortices. This leads to new dissipative effects, such as the vortex mediated mutual friction. The superfluid constraint does, however, drastically simplify the 
dissipation problem and may be a reasonable approximation at temperatures well below the superfluid transition temperature. The
simplified model may also give some insight into which dissipation channels are likely to make the most important  contributions.

As examples of relevant applications, that take us (quite far) beyond the level that has been previously considered, we outlined two problems where a three-fluid description is appropriate.
In fact, although two-fluid hydrodynamics is a useful approximation for the study of a system of neutrons and protons at low temperature (i.e. the outer core of a neutron star) \citep{monster,fmode}, there will always be regions of the star that are close to the superfluid transition temperature and in which the thermal excitations of the system (the entropy in our language) should be considered as a separate component \cite{thermal}.
Furthermore, several equations of state predict the appearance  of hyperons in neutron star cores, with $\Lambda$ and $\Sigma^{-}$ expected to have the lowest threshold densities \citep{LO,hyper}.  The effects of multi-fluid hydrodynamics on the bulk viscosity damping of the gravitational-wave driven r-mode instability have  already been studied in \cite{hyperon}, although in that case the model was simplified to the two-fluid level by neglecting the $\Lambda$ hyperons. A ``complete'' model would require a three-fluid description to account for neutrons, $\Lambda$ hyperons and a charge neutral fluid of protons, electrons and $\Sigma^{-}$ hyperons.

Another interesting possibility is that the ground state of matter at the extreme densities of neutron star cores may correspond to deconfined quarks. At asymptotically high densities matter is expected to be in the CFL state, and one can thus model the system by considering a two-fluid model of a CFL condensate coupled to phonons (the entropy). For realistic core densities it is, however, believed that the ground state will be represented by the so-called CFL-K$^0$ phase \citep{alf1} in which one also has a kaon condensate. Motivated by this, we presented an example of a three-fluid system given by two superfluids, the CFL condensate and the kaons, coupled to a phonon gas. This example is interesting also because it can easily be adapted to describe neutrons and protons coupled to thermal excitations by replacing the kaon and CFL densities with the neutron and proton densities. The two systems are formally equivalent.

From a technical point-of-view, the model developed in this paper takes us to the level where we need to focus on more detailed applications. In order to do this, we must give some thought to the nature of the many new dissipation coefficients that the system allows. Are there situations where the additional channels are important? 
Our experience from neutron star oscillations and the case of the superfluid mutual friction suggests that the answer 
is non-trivial (as one would expect), and that one may have to work out the detailed dynamics of each model
system before knowing for certain. A better insight into the relative strength of the 
different dissipation coefficient and their nature (e.g. the scaling with the key variables of the problem, like density and temperature) would be very useful, and may suggest suitable simplifications for any problem at hand. 
Of course, this presents us with a challenge given that most considerations of NS transport phenomena are
based on the single-fluid model. We  need to move beyond this level to make further progress.

\section*{Acknowledgments}

We wish to thank Mark Alford, Simin Mahmoodifar, and Kai Schwenzer for useful
discussions.
NA acknowledges support from STFC in the UK via grant number ST/J00135X/1. BH acknowledges support from a  EU Marie-Curie Intra-European Fellowship, project number 252470, AMXP dynamics. GLC acknowledges partial support from NSF via grant number
PHYS-0855558, and the hospitality of the Department of Physics of Washington
University where part of this work was carried out.
This work was partially supported by CompStar, a Research Training Network Programme of the European Science Foundation.


\begin{thebibliography}{99}

\bibitem{ligo} 
Abbott, B.P. et al (the LIGO-Virgo Scientific Collaboration), Rep. Prog. Phys. 72, 076901 (2009).

\bibitem{etpaper}
Andersson, N., Ferrari, V., Jones, D. I.,  Kokkotas, K. D., Krishnan, B.,  Read, J. S.,  Rezzolla, L., Zink, B., Gen. Rel. Grav. 43, 409 (2011)

\bibitem[\protect\citeauthoryear{Easson \& Pethick}{Easson \& Pethick}{1979}]{EP} Easson I., Pethick C.J., ApJ 227, 995 (1979)

\bibitem[\protect\citeauthoryear{Andersson \& Comer}{Andersson \& Comer}{2007}]{monster} Andersson N., Comer G.L.,  Class. Quantum. Grav., 23, 5505
(2007)

\bibitem{sfr1}
Passamonti, A., Haskell, B., Andersson, N., MNRAS,  396, 951 (2009)

\bibitem[\protect\citeauthoryear{Haskell \& Andersson}{Haskell \& Andersson}{2010}]{hyperon} Haskell B., Andersson N., MNRAS 408, 1897 (2010)

\bibitem[\protect\citeauthoryear{Haskell, Andersson \& Passamonti}{Haskell, Andersson \& Passamonti}{2009}]{rmode} Haskell B., Andersson N., Passamonti A., MNRAS 397, 1464 (2009)

\bibitem[\protect\citeauthoryear{Lindblom \& Mendell}{Lindblom \& Mendell}{1995}]{LM} Lindblom L., Mendell G., ApJ 444, 804 (1995)

\bibitem[\protect\citeauthoryear{Yoshida \& Lee}{Yoshida \& Lee}{2003}]{YL} Yoshida S.,  Lee U., MNRAS 344, 207 (2003)

\bibitem[\protect\citeauthoryear{Gusakov \& Kantor}{Gusakov \& Kantor}{2008}]{Ghyp} Gusakov M.E., Kantor E.M., Phys. Rev. D 78, 83006 (2008)

\bibitem[\protect\citeauthoryear{Lindblom \& Owen}{Lindblom \& Owen}{2002}]{LO} Lindblom L., Owen B.J., Phys. Rev D., 65, 063006 (2002)

\bibitem[\protect\citeauthoryear{Nayyar \& Owen}{Nayyar \& Owen}{2006}]{nayyar} Nayyar B.,  Owen B.J., Phys. Rev. D 73, 084001(2006)

\bibitem[\protect\citeauthoryear{Andersson, Haskell \& Comer}{Andersson, Haskell \& Comer}{2010}]{cfl} Andersson N., Haskell B., Comer G.L., Phys. Rev. D 82, 023007 (2010)

\bibitem{hoprl} 
Ho, W.C.G, Andersson, N., Haskell, B., Phys. Rev. Lett.  107, 101101 (2011)

\bibitem{degen}
Haskell, B, Degenaar, N., Ho, W.C.G., {\em Constraining the physics of the r-mode instability in neutron stars with X-ray and UV observations}, preprint arXiv:1201.2101

\bibitem[\protect\citeauthoryear{Prix}{Prix}{2004}]{Prix} Prix R., Phys. Rev. D, 69, 043001 (2004)

\bibitem{chamel1}
Carter, B., Chamel, N.,  Int. J. Mod. Phys. D 13,  291 (2004)

\bibitem{chamel2}
Carter, B., Chamel, N.,  Int. J. Mod. Phys. D 14,  717 (2005)

\bibitem[\protect\citeauthoryear{Glendenning}{Glendenning}{1985}]{Glenn} Glendenning N.K., ApJ 293, 470 (1984)

\bibitem[\protect\citeauthoryear{Alford, Schmitt, Rajagopal \& Schafer}{Alford et al.}]{reviewA} Alford M.G., Schmitt A., Rajagopal K., Sch\"afer T., Rev. Mod. Phys. 80, 1455 (2008)

\bibitem[\protect\citeauthoryear{Alford, Braby \& Schmitt}{Alford, Braby \& Schmitt}{2008}]{alf1} Alford M.G., Braby M., Schmitt A., J. Phys. G. 35, 115007 (2008)

\bibitem[\protect\citeauthoryear{Alford, Braby, Reddy \& Sch\"{a}fer}{Alford et al.}{2007}]{kaon1} Alford M.G., Braby M., Reddy S., Sch\"{a}fer T., Phys. Rev. C 75, 055209 (2007)

\bibitem{nuclpa}
Andersson, N., Comer, G. L.,  Glampedakis, K., Nucl. Phys. A  763, 212 (2005)

\bibitem{thermal}
Andersson, N., Kr\"uger, C., Comer, G.L., Samuelsson, L., {\em A minimal model for finite temperature superfluid dynamics} in preparation 


\bibitem[\protect\citeauthoryear{Andersson, Comer \& Glampedakis}{Andersson, Comer \& Glampedakis}{2005}]{visc} Andersson N., Comer G.L., Glampedakis K., Nucl. Phys. A 763, 212 (2005)

\bibitem[\protect\citeauthoryear{Cutler \& Lindblom}{Cutler \& Lindblom}{1987}]{visc1} Cutler C., Lindblom L., ApJ 314, 234 (1987)


\bibitem[\protect\citeauthoryear{Kantor \& Gusakov}{Kantor \& Gusakov}{2009}]{kant} Kantor E.M., Gusakov M.E., Phys. Rev. D 79, 43004 (2009)

\bibitem[\protect\citeauthoryear{Andersson \& Comer}{Andersson \& Comer}{2008}]{helium} Andersson N., Comer G.L., Int. J. Mod. Phys. D, 20, 1215 (2011)

\bibitem[\protect\citeauthoryear{Khalatnikov}{Khalatnikov}{1965}]{kal} Khalatnikov I.M, {\em An introduction to the theory of superfluidity} (W.A. Benjamin, New York, 1965)

\bibitem[\protect\citeauthoryear{Andersson \& Comer}{Andersson \& Comer}{2010}]{heat} Andersson N., Comer G.L., Proc. R. Soc. A 466, 1373 (2010)

\bibitem{extro}
Jou, D., Casas-V\'azquez, J., and Lebon, G., {\em Extended irreversible thermodynamics}, (Springer, New York, 2010)

\bibitem[\protect\citeauthoryear{Glendenning}{Glendenning}{2000}]{GlennBook} Glendenning N.K., {\em Compact Stars: Nuclear Physics, Particls Physics and General Relativity}, 2nd Edition (Springer Verlag, New York, 2000)

\bibitem[\protect\citeauthoryear{Glampedakis, Andersson \& Samuelsson}{Glampedakis et al.}{2011}]{supercon} Glampedakis K., Andersson N., Samuelsson L., MNRAS 410, 805 (2011)

\bibitem[\protect\citeauthoryear{Mendell}{Mendell}{1991a}]{Mendella} Mendell G., ApJ., 380, 515 (1991)

\bibitem{carter}
Carter, B., {\em Covariant Theory of Conductivity in Ideal Fluid or Solid Media}, pp 1-64 in Relativistic Fluid Dynamics, Lectures given at the 1st 1987 session of the Centro Internazionale Matematico Estivo (C.I.M.E.) held at Noto, Italy, May 25 Ð June 3, 1987,  Ed. A. Anile \& M. Choquet-Bruhat, vol. 1385 of Lecture Notes in Mathematics,  (Springer, Berlin, 1989)

\bibitem[\protect\citeauthoryear{Andersson \& Comer}{Andersson \& Comer}{2006}]{livrev} Andersson N., Comer G.L., Living Rev. Relativity 10, http://www.livingreviews.org/lrr-2007-1(2007)

\bibitem[\protect\citeauthoryear{Comer \& Joynt}{Comer \& Joynt}{2003}]{cj} Comer G.L., Joynt R., Phys. Rev. D 68, 023002 (2003)

\bibitem[\protect\citeauthoryear{Gusakov, Kantor \& Haensel}{Gusakov, Kantor \& Haensel}{2009b}]{Gusakov2} Gusakov M.E., Kantor E.M., Haensel P., Phys. Rev. C 79, 015803 (2009)

\bibitem[\protect\citeauthoryear{Gusakov, Kantor \& Haensel}{Gusakov, Kantor \& Haensel}{2009a}]{Gusakov1} Gusakov M.E., Kantor E.M., Haensel P., Phys. Rev. C 79, 055806 (2009)

\bibitem[\protect\citeauthoryear{Lopez-Monsalvo \& Andersson}{Lopez-Monsalvo \& Andersson}{2010}]{cesar} Lopez-Monsalvo C.S., Andersson N., Proc. R. Soc. London A,  467, 738 (2011)

\bibitem{cesar2}
Andersson, N., and Lopez-Monsalvo, C.S., Class. Quantum Grav. {\bf 28}, 195023 (2011)

\bibitem{mhd}
Andersson, N., {\em Resistive relativistic magnetohydrodynamics from a charged multi-fluids perspective} preprint arXiv:1204.2695

\bibitem[\protect\citeauthoryear{Onsager}{Onsager}{1931}]{Onsager} Onsager L., Phys. Rev.  37, 405 (1931)

\bibitem{prigogine} Kondepudi D, Prigogine I., {\em Modern Thermodynamics} (John Wiley and Sons, Chichester, 2005)

\bibitem{chamel} Carter, B., Chamel, N., Int. J. Mod. Phys. D, Volume, 14,  749 (2005)

\bibitem{megus}
Gusakov, M.E., Phys. Rev. D. {\bf 76}  083001 (2007)

\bibitem{manna}
Manuel, C., Mannarelli, M., Phys. Rev. D {\bf 81} 043002 (2010)


\bibitem[\protect\citeauthoryear{Alpar, Langer \& Sauls}{Alpar et al.}{1984}]{mf2} Alpar M.A., Langer S.A., Sauls J.A., ApJ, 282, 533 (1984)

\bibitem[\protect\citeauthoryear{Mendell}{Mendell}{1991b}]{Mendellb} Mendell G., ApJ., 380, 530 (1991)

\bibitem[\protect\citeauthoryear{Andersson, Sidery \& Comer}{Andersson, Sidery \& Comer}{2006}]{trev} Andersson N., Sidery T.L., Comer G.L., 
MNRAS 368, 162 (2006)

\bibitem[\protect\citeauthoryear{Melatos \& Peralta}{Melatos \& Peralta}{2007}]{Mp1} Melatos A., Peralta C., ApJ 662, L99 (2007)

\bibitem[\protect\citeauthoryear{Melatos \& Peralta}{Melatos \& Peralta}{2010}]{Mp2} Melatos A., Peralta C., ApJ 709, 77 (2010)

\bibitem{trev2}
Andersson, N.; Sidery, T.; Comer, G. L., MNRAS, 381, 747 (2007)

\bibitem[\protect\citeauthoryear{Babaev}{Babaev}{2009}]{babaev} Babaev E., Phys. Rev. Lett. 103, 231101 (2009)

\bibitem{zit}
Buckley, K.B., Metlitski, M.A., Zhitnitsky, A.R., Phys. Rev. C 69, 055803 (2004)

\bibitem{lm1}
Lindblom, L., Mendell, G., Ap. J. 421,  689 (1994)

\bibitem{lm2} 
Lindblom, L., Mendell, G., Phys. Rev. D 61, 104003 (2000)

\bibitem[\protect\citeauthoryear{Andersson, Glampedakis \& Haskell}{Andersson et al.}{2009}]{fmode} Andersson N., Glampedakis K., Haskell B., 
Phys. Rev D 79, 103009 (2009)

\bibitem[\protect\citeauthoryear{Andersson \& Kokkotas}{Andersson \& Kokkotas}{2010}]{NilsRev} Andersson N., Kokkotas K.D, Int. J. Mod. Phys. D 10, 381 (2001)

\bibitem{demorest}
Demorest, P.B., Pennucci, T., Ransom, S.M., Roberts, M.S.E., Hessels, J.W.T., Nature 467,  1081 (2010)

\bibitem{stone}
Stone, J.R., Guichon, P.A.M., Thomas, A.W., {\em Role of Hyperons in Neutron Stars}, preprint arXiv:1012.2919

\bibitem[\protect\citeauthoryear{Haensel, Levenfish \& Yakovlev}{Haensel, Levenfish \& Yakovlev}{2002}]{hyper} Haensel P., Levenfish K.P., Yakovlev D.G., A\&A 381, 1080 (2002)


\bibitem[\protect\citeauthoryear{Rupak \& Jaikumar}{Rupak \& Jaikumar}{2010}]{RJ} Rupak G., Jaikumar P., Phys. Rev. C., 82, 055806 (2010)


\bibitem[\protect\citeauthoryear{Alford, Braby \& Mahmoodifar}{Alford, Braby \& Mahmoodifar}{2010}]{kaon2} Alford M.G., Braby M., Mahmoodifar S., Phys. Rev. C 81, 025202 (2010)

\bibitem{gus}
Gusakov, M.E., Andersson, N., MNRAS 372, 1776 (2006)




\end{thebibliography}
\end{document}